# Supervised Learning for Game Music Segmentation

Shangxuan Luo, Joshua Reiss
Queen Mary University of London

*Abstract*—At present, neural network-based models, including transformers, struggle to generate memorable and readily comprehensible music from unified and repetitive musical material due to a lack of understanding of musical structure. Consequently, these models are rarely employed by the games industry. It is hypothesised by many scholars that the modelling of musical structure may inform models at a higher level, thereby enhancing the quality of music generation. The aim of this study is to explore the performance of supervised learning methods in the task of structural segmentation, which is the initial step in music structure modelling. An audio game music dataset with 309 structural annotations was created to train the proposed method, which combines convolutional neural networks and recurrent neural networks, achieving performance comparable to the state-of-the-art unsupervised learning methods with fewer training resources.

## I. Introduction

Video games have long been regarded as an appropriate medium for music generation, given their interactive and non-linear nature (Collins 2009). Nevertheless, it is evident that the current game industry employs only a limited degree of music generation technology. Worrall and Collins (2023) conducted a study in which they surveyed 11 professional game music composers about their concerns and considerations regarding AI music technology. The most often cited concern was that AI-generated music lacks originality and quality compared to human-written compositions. The inconsistent musical quality and lack of human nuance in the output have the potential to negatively impact the gameplay experience. One example is provided by the audio director of *No Man's Sky*[1], Paul Weir, who notes that generative music was used in the game with an acknowledgement that it could produce "worse" music than composed music. Furthermore, generative music is often unpredictable and can be difficult to control.

In order to enhance the quality of model-generated music, many scholars have identified the modelling of musical structure as an important way for improvement. Jhamtani and BergKirkpatrick (2019) posit that the learning of musical structure enables music generation models to generate recurring, more coherent music. Bhandari and Colton (2024) describes the key guidance of music structure for music generation, they referred it as "sub-task decomposition", which decompose music generation into separate high-level structural planning and content creation stages. The modelling of musical structures enables the capture of recurring themes and melodies, allowing the model to learn the nuances of musical variation in a way analogous to that of human composers. Furthermore, it facilitates the generation of highly unified musical compositions within a defined structural framework.

The majority of games comprise a soundtrack between one and four hours of music, whereas the duration of gameplay can vary considerably, from six to over one hundred hours (Cullimore et al., 2014). This often results in players hearing musical tracks repeated on many occasions. While repetition is a key element in music, an excess of repetition can disrupt immersion and become a source of frustration for the player (Scirea 2017, Sweet 2015). Modelling the structure of game music could also enhance the cost-effectiveness of procedural music generation, empowering composers through assistive means (Plut 2020).

However, due to the hierarchy and subjectivity nature of music, music structure modelling is still an open and under-explored field. The primary challenges include subjectivity, ambiguity, and hierarchy (Nieto et al. 2020). Firstly, subjectivity arises because different people might disagree about the structure of a particular music. Secondly, ambiguity occurs when the same person accepts multiple valid interpretations based on the musical attributes they focus on. Lastly, hierarchy presents a challenge as musical structures exist at multiple timescales simultaneously. This introduces variability in the annotations, making it a hard problem to solve.

Recently, an increasing number of researchers have investigated the potential of data-driven AI techniques for music structure modelling, inspired by the success of these techniques in other fields. Yet compared with other fields, the structured-annotated music data sets are rare, and there is no publicly available game music data set for structure analysis. This study contributes two aspects to the field. First, it provides a new data set with 309 structural annotations, which collects video game music since 2000, completed by two annotators with formal music training backgrounds, including the author. Second, it explores the performance of a neural network-based supervised learning approach on a proposed data set.

## II. Related work

The target of music structure modelling or music structure analysis (MSA) can be dissembled into two tasks: firstly, to determine the boundaries and constituent segments of a given musical piece, and secondly, to identify and compare the similarities between these segments. This paper is concerned with the investigation of the first stage.

Traditional methods extract audio features from music content first to generate a self-similarity Matrix (SSM) and then detect the boundaries and compare similarities within the SSM. The checkerboard kernel technique by Foote (2000) is effective and simple to use by considering local details. This method essentially uses a kernel to probe the change curve of musical novelty, with the peak indicating a potential boundary. However, it is challenging to differentiate between homologous paragraphs with high similarity. In music structure modelling, novelty peaks are commonly observed in musical paragraph boundaries; however, this does not necessarily imply that the opposite is true. Consequently, the method may yield an increased number of false positives when used alone.

---

[1] https://www.gdcvault.com/play/1024067/The-Sound-of-No-Man

Other methods include lag matrices (Goto 2003) and structural features (Serrà et al. 2014), which enhance SSM computation. A lag matrix helps to find repeated patterns in a sequence by comparing each point with a few of its previous points and looking for similar patterns in these comparisons. The idea of structure features is further to consider both local and global details by comparing each frame in a sequence to every other frame. In addition, some post-processing techniques such as path smoothing (Serran et el. 2009), transposition-invariant (Müller and Clausen 2007), threshold strategy are used to improve the recognition accuracy.

Although the traditional method based on SSM requires less computational resources, it requires very careful feature selection and complex parameter fine-tuning. The feature type, the window size and the temporal resolution used for feature extraction crucially determine whether blocks or stripes are formed in an SSM (Müller 2021) tailored to different music genres.

Due to the high dependence of machine learning methods on data volume and the scarce resources of publicly available datasets, there are relatively few studies using these techniques on music structure modelling. Wang et al. (2022) trained a deep neural network model in a supervised way to output an embedding to improve the traditional algorithm. McCallum (2019) introduced an unsupervised method combined Convolutional Neural Network (CNN) with contrastive learning to leverage available unlabelled data. Buisson (2024) continued his research on multi-level audio representations. Both methods achieved better results compared to traditional methods.

Given the success of deep learning method, this paper uses a deep neural network architecture combining CNN and Recurrent Neural Network (RNN) to train on a unified self-built dataset in a supervised way.

### III. METHODOLOGY

#### A. Dataset

Mainstream datasets used for music structure modelling or MSA in audio domain are displayed in Table 1. The Structural Annotations for Large Amounts of Music Information (SALAMI) dataset (Smith et al. 2011) is the largest publicly available dataset with hierarchical annotations for 1,359 tracks, provided by a total of 10 music experts. Out of these, 884 tracks have annotations from two different annotators. The dataset encompasses a diverse selection of music genres: classical, jazz, popular, world, and live music.

The Harmonix Set (Nieto et al. 2019) is focused primarily on popular western music genres, including hip-hop, dance, rock, and metal. It includes annotations for 912 tracks. For each song, a tempo track was created using Digital Audio Workstation software (e.g., Logic Pro), with beats, downbeats, and segments added to it. As a result, the segment boundaries in this dataset consistently align with annotated beats.

The Real World Computing (RWC) dataset, also known as the AIST Annotations (Goto, 2006), includes 300 tracks using single flat structural segments. The music styles covered range from pop to classical, with a substantial jazz subset. Similar to the Harmonix Set, boundaries in this dataset are consistently aligned with beat positions.

Isophonics was originally compiled by the Centre for Digital Music (C4DM) at Queen Mary University of London (Mauch et al., 2009). This dataset consists of 300 tracks of Western popular music, primarily pop-rock, with flat, coarse-level segmentation. It also includes beat and downbeat annotations for The Beatles subset, which can be utilized by algorithms that work at the beat level.

A key advantage of SALAMI is its detailed annotation guidelines, which this paper adapts when creating the proposed dataset for video game music. The annotations are categorized into three hierarchical levels: (i) the fine level, for short motives or phrases, (ii) the coarse level, for larger sections like paragraphs, and (iii) the functional level, which provides semantic labels (e.g., verse, bridge) that overlap with the coarse level. Since game music has no specific functional terms, the author chose the following basic terms that are most similar to the SALAMI annotation guide that is most appropriate for video game music: Intro, Main Theme, Secondary Theme, Transition and Outro. An example can be seen in Figure 1.

TABLE I. COMPARISON WITH EXISTING MUSIC DATASETS WITH STRUCTURAL ANNOTATION

| Name | Year | Tracks | Annotator | Genres | Hierarchy | Beat Synchronization |
|---|---|---|---|---|---|---|
| SALAMI | 2011 | 1359 | 10[a] | Classical, Jazz, Popular, World, Live Music | Yes | No |
| Harmonix | 2019 | 912 | 1 | Hip Hop, Dance, Rock, Metal | No | Yes |
| RWC | 2006 | 300 | 1 | Pop, Classical, Jazz | No | Yes |
| Isophonics | 2009 | 300 | 1[b] | Western Popular Music, primarily Pop-Rock | No | Yes |
| **SAVGM** | **2024** | **309** | **2** | **Video Game Music** | **Yes** | **No** |

[a.] 884 tracks have annotations from two different annotators

[b.] The Beatles part of annotations were initially collected by Alan Pollack, the rest were collected by experts at C4DM.

```
0.000000000      [b], a, I, intro
12.353015873     [t], b, A, main theme
24.488843537     [rp], b'
36.579319728     [t], c, B, secondary theme
48.535918367     [rp], c
60.518072562     [t], c'
72.510816327     [rp], c'
84.563424036     [t, rp], b, A, main theme
96.558412698     [rp], b'
108.580975057    [t], c, B, secondary theme
120.602312925    [rp], c
132.566802721    [rp], c
144.539206349    [rp], c'
158.824489796    [t, rp], b', A, main theme
171.700294785    [rp], b'
189.611587302    [e]
```

Fig. 1. Example of an annotation file: The first column represents the timestamp. The abbreviations in square brackets indicate the perceived boundary category. Lowercase letters denote annotations at the fine level or music sentence level, while uppercase letters represent annotations at the coarse level or music segment level, with each uppercase letter corresponding to a functional annotation.

The Structural Annotations for Video Game Music (SAVGM) proposed by this paper selected 309 tracks of video game music. Half of these were selected from the top 150 popular music on YouTube[2] for their representativeness, while the remainder were selected by the annotators according to their personal preferences from the top 100 publicly available OST dataset[3] from the 2000s. Following SALAMI's annotation guidelines, three levels of annotation were provided by two annotators with formal music training from the Shanghai Conservatory of Music. The data set does not employ beat or downbeat synchronisation methods, given that video game music typically exhibits greater flexibility in instrument selection and beat patterns, which serve to align with the stylistic and functional nuances of different games. This contrasts with the more fixed instrument configurations and rhythm patterns observed in Western popular music.

Unlike the symbolic domain, the music structure modelling in the audio domain in this paper is more dependent on cognition and perception of musical hearing rather than visual cues. The Gestalt Principles explain how humans naturally perceive and organize visual information by grouping objects and patterns to make sense of complex images. Lerdahl and Jackendoff (1983) applied the Gestalt Principle to the music domain in their book *A Generative Theory of Tonal Music*, which introduced a series of well-formed Grouping Preference Rules (GPR), among which the Rest rule (GPR2a) and Attack Point rule (GPR2b) were proven to be the most effective principles by Frankland (2004). To provide more reference for future investigation of the subjectivity of human annotators, the authors introduced an additional annotation rule based on the GPR to record the main judgment factors of the boundary perceived by the annotators.

The perception of boundaries is informed by two fundamental principles. The Novelty Principle is predicated on the perception of local changes within continuous music. It includes: Rhythm (r): Includes rest and attack-point intervals, as well as rhythm changes; Dynamic (d): Covers variations in intensity, such as accents and downbeats; Timbre (t): A strong cue that includes changes in instruments and articulation; Pitch (p): Encompasses perceived octave changes and transposition; Harmony (h): Relates to changes in chords and harmonic colour.

Similarity Principle considers more global information. Regularity (rg): Indicates regular patterns, with durations of equally annotated segments typically spanning integer ratios of beats and distributed log-normally across the track; Repetition (rp): Reflects melody repetition or highly similar contours. The abbreviations in square brackets indicate the perceived boundary category, as seen in Figure 1.

Furthermore, special labels are included for the purposes of data processing, specifically Begin (b) and End (e). All annotations were conducted using the Sonic Visualiser, a software developed by C4DM at Queen Mary University of London[4].

*B. Prepocessing*

The preprocessing step for feature extraction in audio analysis involves several stages to transform raw audio signals into standardized features.

First, all audio files are resampled to a sampling rate of 22,050 Hz using the Librosa library[5], which provides reliable methods for managing audio data. This resampling is important for ensuring consistent temporal resolution across different files, making the extracted features comparable. Three types of features are extracted from the audio signal, each highlighting different aspects of the sound:

- Mel-Frequency Cepstral Coefficients (MFCCs) are computed using a window size of 2048 and a hop size of 512, as well as a mel scale of 13, to capture the short-term power spectrum of the audio signal on a mel scale, which is useful for representing the timbral quality of sound.

- Constant-Q Transform (CQT) provides a frequency representation of the audio signal with logarithmic frequency resolution. It is used for analyzing harmonic and tonal elements of music in recent literature (Nieto and Bello 2016). The CQT magnitude is computed to focus on the amplitude of frequency components.

- Onset Envelope is used instead of beats or downbeats synchronization to capture the temporal dynamics of the audio signal, highlighting moments of significant energy change that often indicate the onset of new audio events.

In order to guarantee that all features are on an equal scale and to enhance the model's generalisability, each audio feature is normalised using Z-score normalisation. This is achieved by first subtracting the mean and then dividing by the standard deviation for each audio feature. It helps to standardize varying signal amplitudes and ensures that each feature contributes equally during the learning process.

Finally, the normalized features are concatenated along the feature axis, creating a composite feature set that integrates various dimensions of the audio signal. This combined feature

---

[2] https://youtube.com/playlist?list=PLJCXrsW_esBlb1BYfeNhY4kJQuNYoDpF_&si=FAmgBzIg3Ic_vXok
[3] https://downloads.khinsider.com/all-time-top-100
[4] https://www.sonicvisualiser.org
[5] https://github.com/librosa/librosa

vector is then prepared for the subsequent stages of the audio analysis pipeline.

*C. Model*

The most effective boundary retrieval algorithms in recent research are based on deep convolutional architectures (Grill and Schlüter 2015, Ullrich et al. 2014). McCallum's recent work (2019) achieved state-of-the-art results by learning audio features through unsupervised representation learning with convolutional neural networks optimized using a triplet loss.

In line with this, the proposed boundary detection model employs CNNs to extract feature maps, complemented by Long Short-Term Memory (LSTM) networks, which represent a variation of RNNs, to capture long-term dependencies, as shown in Figure 2. This designed is to capture local and global information simultaneously in audio data, focusing on binary classification tasks that indicate the presence or absence of a boundary at each time frame.

The CNN module is designed to extract spatial features from the input data. It includes two convolutional layers: the first layer uses a 3x3 kernel size with a stride of 1, a ReLU activation function, and 32 filters, while the second layer uses 64 filters. Padding is set to 1 to maintain spatial dimensions. An ablation experiment showed that adding a Max Pooling layer increased training speed but decreased model performance, so it was not incorporated into the final model. The CNN module outputs a feature map that retains the temporal dimension while enriching the data with extracted features.

The LSTM module processes this output to capture temporal dependencies within the sequence. It consists of a bidirectional LSTM network that considers both past and future contexts. The LSTM has a hidden size of 128 and is stacked with 2 layers to enhance model capacity. To prevent overfitting, a dropout rate of 0.5 is applied between LSTM layers, resulting in an output size of 256 due to the bidirectional nature. The output of LSTM is fed into a fully connected (FC) layer that reduces the dimensionality to a single output per time frame for binary classification (boundary vs. non-boundary).

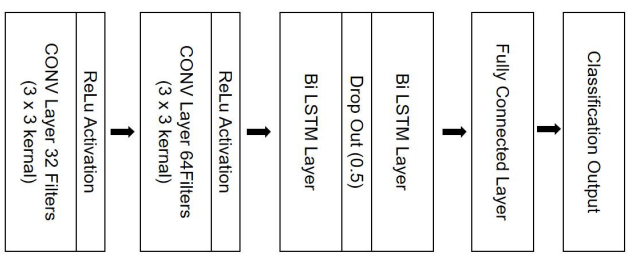

Fig. 2. The architecture of the proposed model

*D. Implementation*

To improve the model's generalization, pre-training is performed using the SALAMI Internet Archive (SALAMI - IA) dataset, which is a publicly accessible subset of the original data set, with only 342 tracks available due to copyright restrictions. The SALAMI-IA is particularly noteworthy because it mainly consists of live recordings with various imperfections (dialogue and background noise, etc.), so it may cause a decline in performance in training alone compared to other data sets. Furthermore, data augmentation techniques, including tempo stretching (0.8 to 1.2 times) and pitch transposition (within 2 semitones), were applied to the training set to enhance the model's performance.

The input data features were set to 98, representing a stack of combined features extracted from the original audio. After pre-training on SALAMI-IA, the model continues training on an augmented training set for 60 epochs using binary cross-entropy with logits. This loss combines a sigmoid layer and binary cross-entropy loss into a single class. To address class imbalance, a positive class weight is set to 100, as the boundary class is significantly outnumbered by the non-boundary class in our case.

The model is optimized using Adam with a learning rate of 0.01. All hyperparameters are tuned on a subset using grid search. Implemented in PyTorch and trained on an A100 GPU, each epoch takes approximately 18 minutes. The model's performance is monitored during training, and the final model is saved for future inference or fine-tuning.

This architecture combines the strengths of CNNs for feature extraction and LSTMs for capturing temporal dependencies, making it well-suited for boundary detection tasks in sequential data. The introduction of a positive class weight effectively mitigates class imbalance, ensuring the model remains sensitive to boundary instances despite their rarity.

*E. Results*

Evaluating the effectiveness of a music boundary classification model commonly involves metrics based on precision, recall, and the F-measure, also referred to as the Hit Rate measure by Turnbull et al. (2007).

In this context, the annotations are binary sequences that represent the presence or absence of music boundaries. The original logits produced by the model are transformed into binary values through a local maximum filter, as demonstrated in Figure 3, to compare with the true annotations. It is important to highlight that retaining the soft probability output is still beneficial, considering the subjective nature of music annotation.

Precision is calculated as the ratio of true positives to the total positive predictions, reflecting how accurately the model identifies positive boundaries. Recall, on the other hand, is defined as the number of true positives divided by the total number of positive boundaries. Precision and recall are often combined by computing their harmonic mean, commonly referred to as the F-measure, which is a metric that seeks to balance the trade-off between precision and recall.

The performance of the model is shown in Table 2, with a 3-second tolerance of boundary match for all assessments (Levy and Sandler, 2008). The unsupervised method by McCallum (2019) serves as the baseline for our assessment. The proposed model achieves competitive performance and requires less computational resources for training compared to the unsupervised methods, which often train on data sets with at least 10,000 tracks.

Despite the fact that the beats synchronisation method has been shown to produce the most favourable results in many studies (Pauwels et al., 2013; McCallum, 2019; Buisson et al., 2024), it did not yield superior outcomes in this paper. One potential explanation is that game music exhibits less intrinsic cohesion compared to other musical styles. The genre of music is often aligned with the style of the game, with notable distinctions between the various genres. Additionally, the structure of music is subject to alteration depending on its function within the game. For instance, theme music tends to exhibit greater irregularity and drama, while background music often displays the opposite characteristics. Enhancing the dataset through an expansion of its size and a more comprehensive representation of the musical styles within it could potentially lead to an improvement in the model's performance.

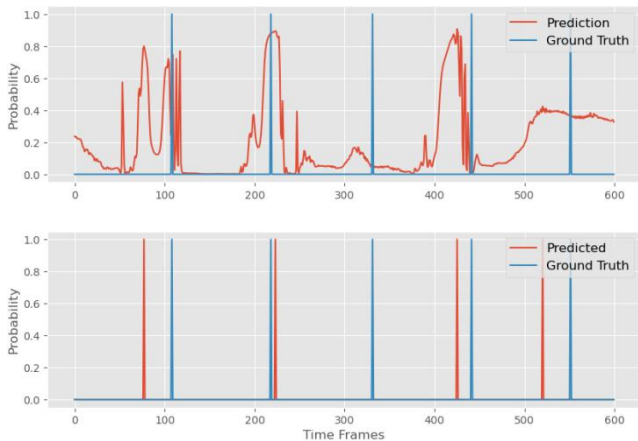

Fig. 3. An example of prediction for the song 'Drink Up, There's More!'. The upper figure is the original logits of the output, and the lower figure is the binary prediction of the output using the local maximum value.

TABLE II. PERFORMANCE METRICS

| Dataset | Metrics | | |
|---|---|---|---|
| | *Precision* | *Recall* | *F1* |
| SALAMI-IA[a] (Unsynchronized) | 0.429 (**0.18**) | 0.653 (**0.15**) | 0.497 (0.16) |
| SALAMI-IA[a] (Beat-Synchronized) | 0.491 (0.20) | 0.660 (0.16) | 0.535 (**0.15**) |
| SALAMI-IA[b] | 0.507 (0.21) | 0.509 (0.19) | 0.477 (0.19) |
| VGM309[b] (Non Pre-trained) | 0.489 (0.33) | 0.589 (0.17) | 0.490 (0.24) |
| VGM309[b] (Pre-trained) | **0.512** (0.32) | **0.667** (0.16) | **0.537** (0.23) |

[a]·Baseline method by McCallum (2019)

[b]·Supervised learning proposed by this paper

## IV. DISCUSSION

When studying music structure modelling problems, the effectiveness of an algorithm is only part of the equation. The inherent limitations of human-labeled datasets, while valuable, can introduce biases, inconsistencies, and errors due to subjective interpretation or varying domain expertise. These limitations can affect the performance evaluation of music structure modelling algorithms, as the "ground truth" provided by human labels may not be entirely accurate or consistent. The primary challenges in music structure modelling are defined as subjectivity, ambiguity, and hierarchy (Nieto et al., 2020). The discussion will address the aforementioned challenges and propose potential solutions with regard to the construction of new game music datasets.

### A. Subjectivity

A significant study on the subjectivity of segment boundary retrieval was conducted by Wang et al. (2017). They approached the issue by crowdsourcing it to a large group of annotators and identified notable differences between strong and weak boundaries, gradual and sudden boundaries, as well as perceptual variations. Empirical evidence suggests a correlation between the inconsistent location of musical boundaries and structural changes. For example, the prelude, foreshadowing and pre-chorus preceding the start of a music sentence or paragraph, in addition to the extension, reproduction and supplementation of the end of a music sentence or paragraph, may be considered. At present, there is no consensus on how the location of the boundary responds to these changes, and there is a paucity of relevant research. The placement of the boundary, whether before a prelude or the beginning of the theme, appears to be a reasonable approach, though not an absolute necessity.

From the perspective of auditory perception, the perception of boundaries is also related to the global arrangement of music. When a prelude overlaps with the theme phrase, it is more susceptible to being subsumed by the prevailing rhythm. Consequently, the perception of boundaries is more likely to emerge before the theme than before the prelude, as shown in Figure 4. Furthermore, these preludes or endings can often be expanded into complete phrases or paragraphs, in accordance with the composer's intentions. The necessity to distinguish between them according to the specific musical context introduces a significant challenge to the establishment of uniform labelling rules.

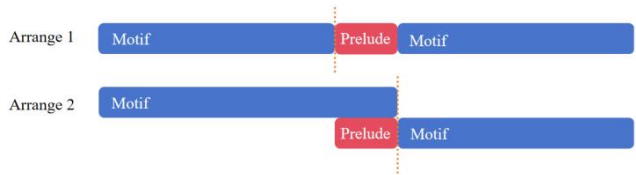

Fig. 4. Different boundary perception by different arrangements. The red dotted line implies a potential boundary.

Despite the provision of explicit annotation guidance, such as the annotation method offered by SALAMI, significant discrepancies remain when data sets are annotated by multiple annotators. This study considers the labels assigned to the same musical piece by different annotators to be distinct works for the purpose of learning. An alternative approach would be to merge the multiple annotations into a single 'gold standard', as proposed by Nieto (2015). While it is challenging to eliminate all subjective differences in annotation, the annotation rule of the proposed data set determines that short structural extensions are divided into main phrases to ensure their integrity. Furthermore, the annotations are mutually verified by two annotators, and the difference boundary is adjusted by consensus.

### B. Ambiguity

A further challenge is presented by the ambiguity of paragraph annotations, which often require the consideration of musical elements from different dimensions. The available evidence indicates that an emphasis on different features can impact the perception of music structure (Smith, 2014). When evaluating the overall similarity of two paragraphs, it is necessary to take into account multiple aspects, including the melody, harmony, texture, timbre, and other dimensions of information simultaneously. This comprehensive understanding makes it challenging to quantify using a computational method.

Furthermore, the degree of global contrast of the phrase will influence the perception of the annotator. When the overall difference degree of music is high, such as in the case of theme music or character music, two similar musical phrases are more likely to be annotated as homologous materials, for example, a and a'. Conversely, when the overall difference degree of music is low, for say, background music or ambient music, two equally similar musical phrases may be annotated as different materials, for example, a and b. One instance from the proposed data set is *Blow*, which uses a consistent thematic rhythm and the contrast between materials is not pronounced, which increases the likelihood of similar material being designated as distinct. This clustering principle can be compared to the visual distinction of colours. If there are only similar colours in a picture, people tend to distinguish them into different colour categories. However, if other contrasting colours appear in the picture at the same time, people are more likely to group similar colours into a colour scheme.

A third point of consideration is that the existing framework for structuring features is not applicable to all forms of game music. For example, the composer of *An Endless Desert* uses a more flexible approach to organizing the musical material, with the initial section of the composition exhibiting a phrase structure of 'aaba cdee fgcb'. The use of theme-based annotated terms does not readily lend itself to actual narrative musical expression or can be fitted to traditional musical forms, such as binary form or ternary form. This has significantly elevated the ambiguity of the annotation.

*C. Hierarchy*

The majority of data sets that are applicable to multi-level annotation, in fact, only provide two levels: fine-level at the phrase level and coarse-level at the phrase level. It should be noted that there is an overlap between the functional level and the coarse-level. Single-level algorithms typically employ coarse-level annotations. However, during the actual annotation process, the author discovered that the number of levels required for different types of music varied. Some music could be extracted into multiple levels, while others could be adequately described by a single level.

Two factors are involved in this inconsistency. One such factor is the duration of the music. Some modern games employ dynamic music systems that adapt the music in real-time based on the player's movements, emotional state, or environmental context. The music therefore is typically divided into discrete segments, which may comprise multiple brief pieces that are integrated to align with the specific context of the game. However, game music that can be learned by the model and manually annotated is static, resulting in the inclusion of short game music with simple structures in the data set. A further factor is the redecomposability of musical phrases, whereby some phrases can be further decomposed into finer levels. As shown in Figure 5, different annotators may provide valid annotations at different levels. In the data set proposed in this paper, annotation guidance explicitly provides coarser annotation levels if multiple fine-level divisions are available.

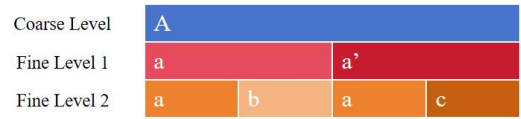

Fig. 5. Redecomposable phrases bring more valid layers

In polyphonic music with multiple melodic lines, it is not uncommon for there to be more than one melodic line. In such cases, two or more musical materials often intersect or overlap, creating the subjective uncertainty of using a single label to label phrases. If a melodic part is dominant, for instance, in a higher pitch, louder volume, or more prominent timbre, the annotation principle outlined in this paper often selects the dominant melody as the reference. When more complex situations arise, such as two materials are in an equal position, it is often challenging to determine which material is the dominant material.

At the time of writing, no model has yet reached the level of performance that would be considered human-equivalent for MSA tasks. Furthermore, the aforementioned challenges in the construction of data sets also have an impact on the construction of models. While the consistency of the annotation helps the model to learn consistent expression, providing large-scale unified annotation data is a challenging task that requires significant time and human resources. It is therefore essential that the designer of the model has a deeper understanding of the annotation principles and musical style of the data set in order to further improve the model performance.

V. FUTURE WORK

This paper examines the application of supervised learning in the context of music structure segmentation on SAVGM. Further research should concentrate on two areas: the combination of segmentation and similarity comparison through the extraction of theme phrases; and the enhancement of the annotation method and the expansion of the data set.

*A. Theme Detection and Extraction*

Improvements in model performance may be achieved in the future through the development of techniques for theme detection and extraction. The perceived categories of boundary changes counted in SAVGM emphasize the importance of repeated phrases for structural modelling, as shown in Figure 6, more than half of perceptual judgments are identified by repetition. Potential difficulties may include the identification of variations of the theme and the extraction of the inner part theme.

Even for those with considerable musical expertise, the identification of certain thematic variations, such as those involving reflections and retrograde motion, can prove challenging. In musical composition, reflection is a technique whereby the melody is altered by symmetrically flipping each note along a fixed axis. For example, if a melody ascends from C to E, the subsequent reflection will descend from C to A. Reflections can alter a melody's fundamental characteristics while retaining some semblance of its original form. This technique is often used by composers to enhance the harmonic and structural aspects of musical compositions. Retrograde is a musical technique whereby a melody or theme is played in reverse. In particular, the notes are rearranged in a reverse order. For example, if the sequence of notes in the original melody was C-D-E-F, the sequence resulting from retrograde would be F-E-D-C. The retrograde technique can imbue musical compositions with a sense of symmetry and logical coherence, and is often employed in fugue and dodecone music.

Furthermore, the alteration of the musical theme between the various parts also presents a challenge in terms of theme extraction. While the majority of the musical themes will initially manifest in the higher part, in order to enhance the musical texture, when the theme reappears, it probably occurs in other parts or is performed by other instruments. At this juncture, the rhythm, interval relationship and harmonic background of the theme may undergo a transformation, and the theme may be obscured by the intrusion of other melodies and harmonies.

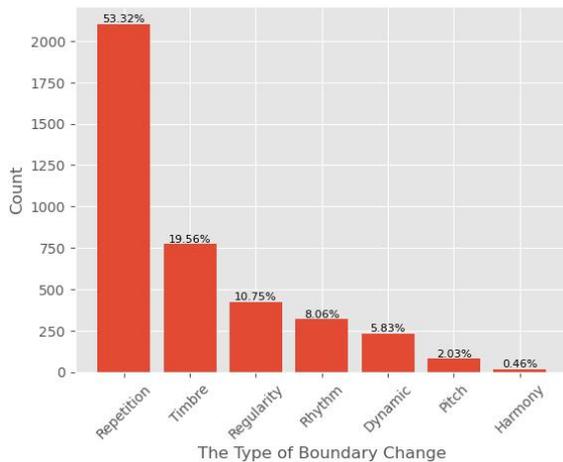

Fig. 6. The Distribution of Types of Boundary Change.

*B. Collect More Annotation Data*

Scaling high-quality data sets is an ongoing process. There is a pressing need for more universally applicable annotation guidance to eliminate the subjective differences among annotators. This guidance should take into account three key elements: it should provide more explicit guidance on the characteristics of strong and weak boundaries, as well as gradual and sudden boundaries. Additionally, it should not be overly reliant on the degree of musical training. Lerdahl and Jackendoff (1983) put forth the proposition that the Rest rule (GPR2a) and Attack Point rule (GPR2b), despite being initially formulated in the context of monophonic music, retain their relevance as fundamental annotation rules in the domain of polyphonic music. Lastly, it should be deemed appropriate for the structural expression of varying game music.

Moreover, the use of multiple annotators and the measurement of inter-annotator agreement, for instance, through the application of Cohen's Kappa, a statistical measure employed to evaluate the level of agreement between two annotators, should be considered as a potential future improvement. A high level of agreement indicates that the annotations are reliable. The combination of annotations from multiple annotators can result in the creation of a consensus label, which can serve to reduce individual biases. Furthermore, the implementation of a review process, whereby annotations are checked for consistency and accuracy, could be adopted to facilitate continuous improvement in annotation quality.